\documentclass[%
reprint,
superscriptaddress,
 amsmath,amssymb,
 aps,
prd,
]{revtex4-2}
\usepackage{graphicx}
\usepackage{dcolumn}
\usepackage{bm}
\usepackage{xcolor}
\usepackage{xspace}
\usepackage{subfigure}
\usepackage{topcapt}

\newcommand{\expNpe}{\ensuremath{\langle\text{N}_\text{PE}\rangle}\xspace}
\newcommand{\Npe}{\ensuremath{n_\text{pe}}\xspace}
\newcommand{\pt}{\ensuremath{p_\text{T}}\xspace}
\newcommand{\mCP}{\ensuremath{\chi}\xspace}
\newcommand{\mCPp}{\ensuremath{\mCP^+}\xspace}
\newcommand{\mCPm}{\ensuremath{\mCP^-}\xspace}

\begin{document}


\title{Search for millicharged particles in proton-proton collisions at $\sqrt{s} = 13$ TeV}

%
%
\author{A.~Ball}\affiliation{CERN, Geneva, Switzerland}
\author{G.~Beauregard}\affiliation{New York University, New York, New York 10012, USA}
\author{J.~Brooke}\affiliation{University of Bristol, Bristol, United Kingdom}
\author{C.~Campagnari}\affiliation{University of California, Santa Barbara,  California 93106, USA}
\author{M.~Carrigan}\affiliation{The Ohio State University, Columbus, Ohio 43218, USA}
\author{M.~Citron}\affiliation{University of California, Santa Barbara,  California 93106, USA}
\author{J.~De~La~Haye}\affiliation{CERN, Geneva, Switzerland}
\author{A.~De~Roeck}\affiliation{CERN, Geneva, Switzerland}
\author{Y.~Elskens}\affiliation{Vrije Universiteit Brussel, Brussel, Belgium}
\author{R.~Escobar~Franco}\affiliation{University of California, Santa Barbara,  California 93106, USA}
\author{M.~Ezeldine}\affiliation{Lebanese University, Hadeth-Beirut, Lebanon}
\author{B.~Francis}\affiliation{The Ohio State University, Columbus, Ohio 43218, USA}
\author{M.~Gastal}\affiliation{CERN, Geneva, Switzerland}
\author{M.~Ghimire}\affiliation{New York University, New York, New York 10012, USA}
\author{J.~Goldstein}\affiliation{University of Bristol, Bristol, United Kingdom}
\author{F.~Golf}\affiliation{University of Nebraska, Lincoln, Nebraska 68588, USA}
\author{J.~Guiang}\thanks{Now at University of California, San Diego, California 92093, USA.}\affiliation{University of California, Santa Barbara,  California 93106, USA}
\author{A.~Haas}\affiliation{New York University, New York, New York 10012, USA}
\author{R.~Heller}\thanks{Now at Fermi National Accelerator Laboratory, Batavia, Illinois 60510, USA.}\affiliation{University of California, Santa Barbara,  California 93106, USA}
\author{C.S.~Hill}\affiliation{The Ohio State University, Columbus, Ohio 43218, USA}
\author{L.~Lavezzo}\affiliation{The Ohio State University, Columbus, Ohio 43218, USA}
\author{R.~Loos}\affiliation{CERN, Geneva, Switzerland}
\author{S.~Lowette}\affiliation{Vrije Universiteit Brussel, Brussel, Belgium}
\author{G.~Magill}\affiliation{McMaster University, Hamilton, Canada}\affiliation{Perimeter Institute for Theoretical Physics, Waterloo, Canada}
\author{B.~Manley}\affiliation{The Ohio State University, Columbus, Ohio 43218, USA}
\author{B.~Marsh}\affiliation{University of California, Santa Barbara,  California 93106, USA}
\author{D.W.~Miller}\affiliation{University of Chicago, Chicago, Illinois 60637, USA}
\author{B.~Odegard}\affiliation{University of California, Santa Barbara,  California 93106, USA}
\author{F.R.~Saab}\affiliation{Lebanese University, Hadeth-Beirut, Lebanon}
\author{J.~Sahili}\affiliation{Lebanese University, Hadeth-Beirut, Lebanon}
\author{R.~Schmitz}\affiliation{University of California, Santa Barbara,  California 93106, USA}
\author{F.~Setti}\affiliation{University of California, Santa Barbara,  California 93106, USA}
\author{H.~Shakeshaft}\affiliation{CERN, Geneva, Switzerland}
\author{D.~Stuart}\affiliation{University of California, Santa Barbara,  California 93106, USA}
\author{M.~Swiatlowski}\thanks{Now at TRIUMF, Vancouver, Canada.}\affiliation{University of Chicago, Chicago, Illinois 60637, USA}
\author{J.~Yoo}\thanks{Now at Korea University, Seoul, South Korea.}\affiliation{University of California, Santa Barbara,  California 93106, USA}
\author{H.~Zaraket}\affiliation{Lebanese University, Hadeth-Beirut, Lebanon}
\author{H.~Zheng}\affiliation{University of Chicago, Chicago, Illinois 60637, USA}

\date{\today}

\begin{abstract}
\noindent We report on a search for elementary particles with charges much smaller than the electron charge using a data sample of proton-proton collisions provided by the CERN Large Hadron Collider in 2018, corresponding to an integrated luminosity of 37.5~fb$^{-1}$ at a center-of-mass energy of 13~TeV.  
A prototype scintillator-based detector is deployed to conduct the first search at a hadron collider sensitive to particles with charges ${\leq}0.1e$. The existence of new particles with masses between 20 and 4700~MeV is excluded at 95\% confidence level for charges between $0.006e$ and $0.3e$, depending on their mass. New sensitivity is achieved for masses larger than $700$~MeV.
\end{abstract}

\maketitle


\section{\label{sec:intro}Introduction}


Over a quarter of the mass-energy of the Universe is widely thought to be some kind of nonluminous ``dark" matter (DM), however, all experiments to date have failed to confirm its existence as a particle, much less its properties. The possibility that DM is not a single particle, but rather a diverse set of particles with as complex a structure in their sector as normal matter, has grown in prominence in the past decade, beginning with attempts to explain observations in high-energy astrophysics experiments~\cite{ArkaniHamed:2008qn, Pospelov:2008jd}.   

Many experimental efforts have been launched to look for signs of a dark sector, including searches at high-energy colliders, explorations at low-energy colliders, precision tests, and effects in DM direct detection experiments (for recent reviews see Refs.~\cite{Battaglieri:2017aum, Beacham:2019nyx,Strategy:2019vxc}).
Most of these experiments target the dark sector via a massive dark photon, in what we refer to as the ``Okun phase"~\cite{Okun:1982xi,Izaguirre:2015eya}. An alternative assumption, which we call the ``Holdom phase''~\cite{Holdom:1985ag, Izaguirre:2015eya}, results in massless dark photons. In these models the principal physical effect is that new dark sector particles that couple to the dark photon will have a small effective electric charge. These are generically called millicharged particles since a natural value for their electric charge of $Q \sim \alpha e/ \pi$ arises from one-loop effects~\cite{Davidson:1993sj}. In this paper we use the symbol \mCP to denote a millicharged particle. For a given mass and charge, the pair production of millicharged particles at the CERN Large Hadron Collider (LHC) is almost model independent. Every standard model (SM) process that results in dilepton pairs through a virtual photon would, if kinematically allowed, also produce $\mCPp\mCPm$ pairs with a cross section reduced by a factor of $(Q/e)^2$ and by mass-dependent factors that are well understood. Millicharged particles can also be produced through Z boson couplings that depend on their hypercharge~\cite{Izaguirre:2015eya}. 
\begin{figure*}[ht]
    \includegraphics[width=7in]{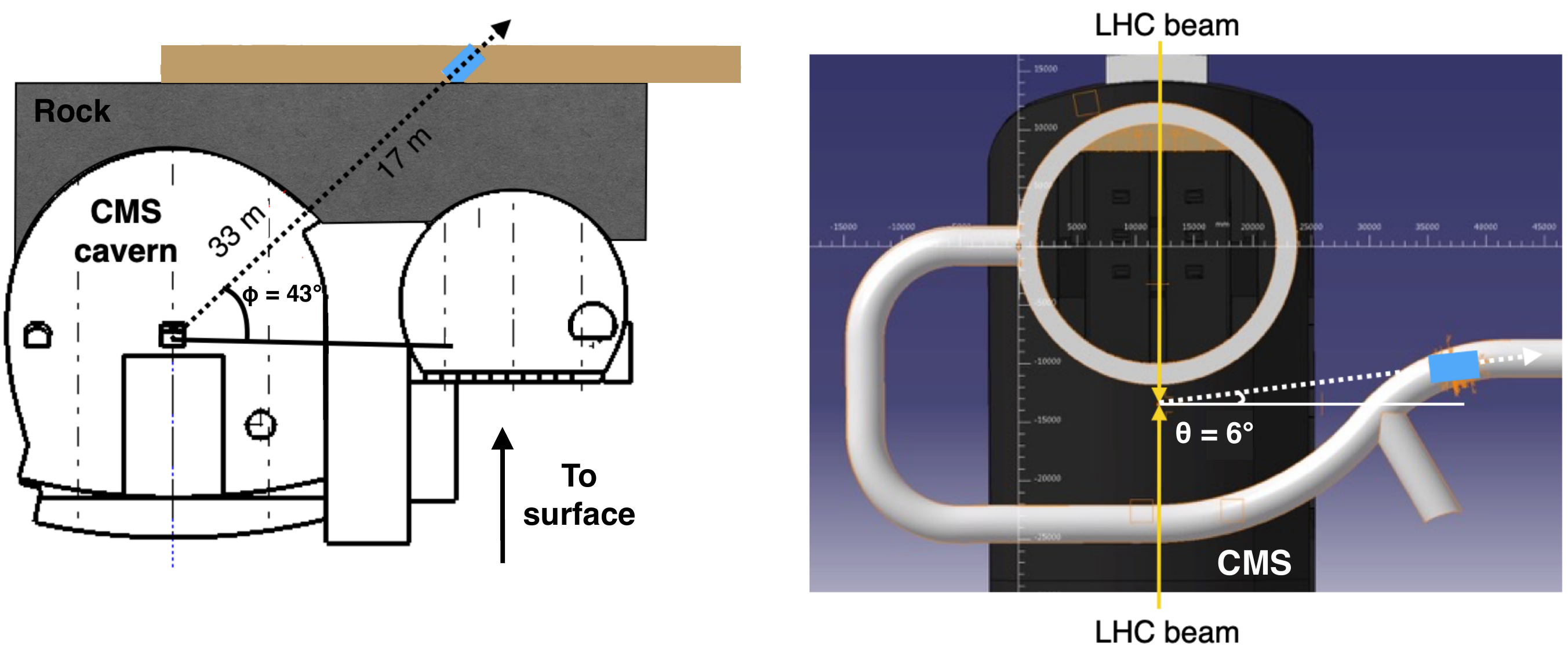}
    \caption{\protect The position of the detector, shown as a blue rectangular volume, in an elevation view (left) and plan view (right).
    The dashed lines represent the projection of the center of the detector to the CMS IP.}
    \label{fig:detectorPosition}
\end{figure*}

Previous experiments have searched for millicharged particles~\cite{MilliQ,essig2013dark,Chatrchyan_2013,Chatrchyan_2013_2,Acciarri_2020, Davidson:2000hf, Badertscher:2006fm}. The parameter space spanned by the mass and charge of \mCP is also constrained by indirect observations from astrophysical systems~\cite{Davidson:1991si, Mohapatra:1990vq, Davidson:1993sj, Davidson:2000hf,Agnese_2015,Emken_2019}, the cosmic microwave background (CMB) ~\cite{Brust:2013ova}, big-bang nucleosynthesis~\cite{Vogel_2014}, and universe overclosure bounds~\cite{Davidson:1991si}. While direct searches robustly constrain the parameter space of millicharged particles, indirect observations can be evaded by adding extra degrees of freedom, which can readily occur in minimally extended dark sector models~\cite{Izaguirre:2015eya}. In particular, the parameter space 1 $ < m_\mCP < 100$~GeV, an ideal mass range for production at the LHC, is largely unexplored by direct (or indirect) searches. Such a signature would not be detectable by the CMS and ATLAS experiments at the LHC ~\cite{ATLASTDR,CMSTDR}, as all detector elements rely on the electromagnetic (EM) interaction of the millicharged particle with ordinary matter. For a millicharged particle the interaction strength is reduced by a factor of $(Q/e)^2$ with respect to that of a particle of the same mass that has charge $e$. The detector signal is also reduced by the same factor, and is typically too small to be recorded by detectors designed for particles of charge $e$. The production of millicharged particles in collider experiments would result in events with missing transverse momentum, however, SM processes with neutrinos as well as instrumental effects tend to overwhelm their signatures. No searches for particles with $Q \lesssim 0.1e$ have been performed at hadron colliders.

It is then clear that dedicated detectors are needed to search for millicharged particles at a hadron collider. In 2016, we discussed the possibility to build such a detector at the LHC, which we called milliQan~\cite{Ball:2016zrp}, at the CMS experimental site and aligned with the CMS interaction point (IP). Since then we have installed and operated a small fraction of such a detector (``milliQan demonstrator'') to measure backgrounds and provide a proof of principle and feedback for the full detector design. In 2018, the demonstrator collected a data set of proton-proton (pp) interactions corresponding to an integrated luminosity of 37.5 fb$^{-1}$,
at a center-of-mass energy of 13 TeV. This corresponds to 86\% of the total luminosity delivered by the LHC in the period the demonstrator was operational.
While the demonstrator is
only $\sim$1\% of the full milliQan, the data collected
already provides competitive constraints on the existence of millicharged particles
of mass 20--4700 MeV$/c^2$ and $Q/e \sim$ 0.01--0.3.

\section{\label{sec:detector}Detector}
A thick sensitive volume is required to be capable of observing the small energy deposition of a particle with $Q \lesssim 0.1e$. The milliQan demonstrator is, therefore, composed of three layers of $80 \times 5 \times 5$ cm  scintillator bar arrays pointing to the CMS IP, with each array consisting of three pairs of bars, stacked on a 3.6~m long rectangular aluminum tube, for a total of 18 bars. We label the closest, middle and furthest layer from the CMS IP as layer~1, layer~2 and layer~3, respectively. 

The milliQan demonstrator is located in an underground tunnel at a distance of 33~m from the CMS IP, with 17~m of rock between the CMS IP and the demonstrator that provides shielding from most particles produced in LHC collisions. In the CMS coordinate system~\cite{CMSTDR}, the detector is positioned at an azimuthal angle ($\phi$) of $43^\circ$ and pseudorapidity ($\eta$) of $0.1$.
Diagrams of the detector's position are shown in Fig.~\ref{fig:detectorPosition}. 
Located 70~m underground, the muon flux from cosmic rays is reduced by a factor of $\sim$100 compared to the surface. 
The detector is aligned using standard laser-based survey
techniques such that the center of the scintillator array projects a line to within 1~cm of the CMS IP. This alignment is validated using
muons produced at the CMS IP, as discussed in Section~\ref{sec:valid}.

\begin{figure}[htp]
    \includegraphics[width=\columnwidth]{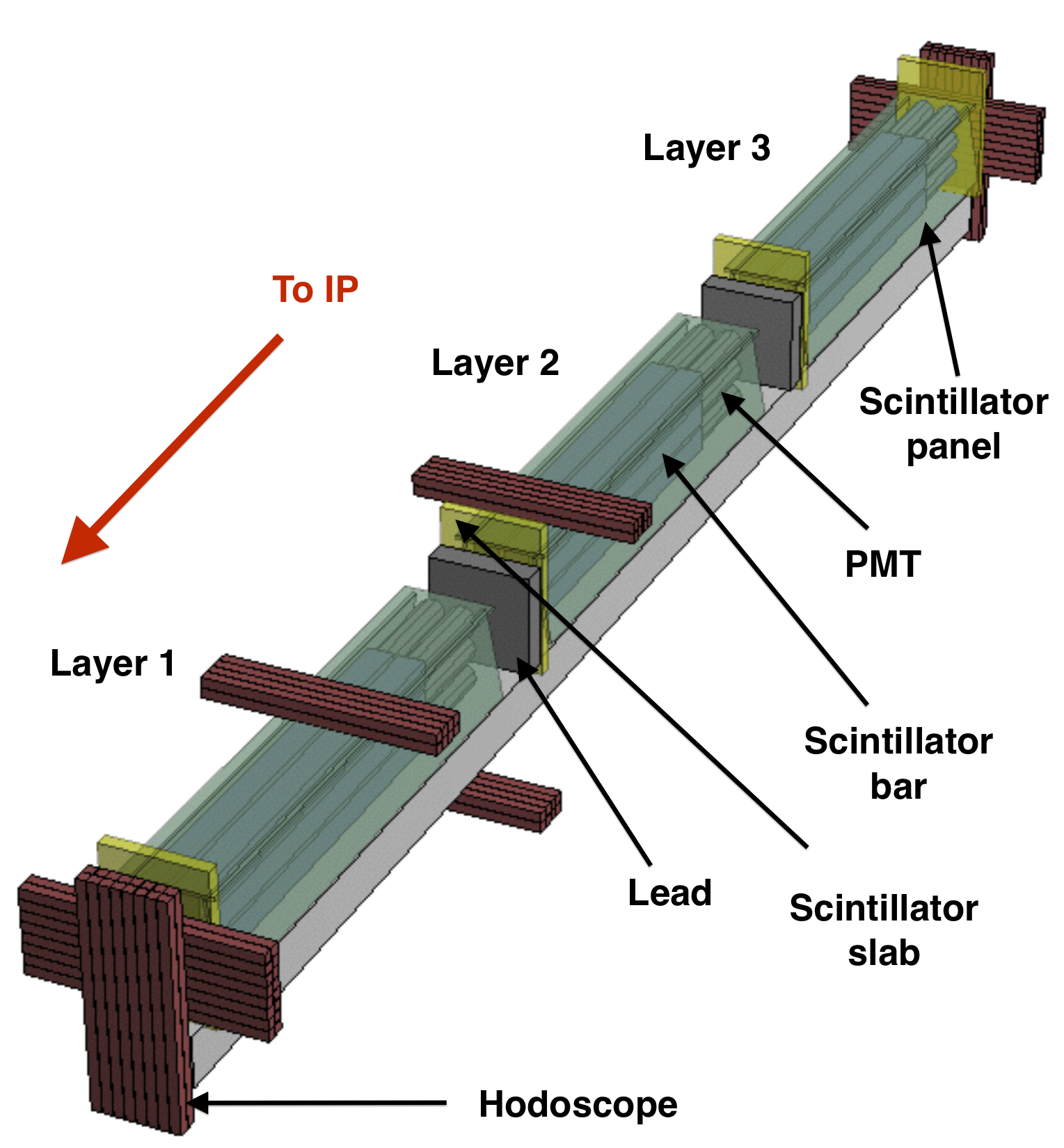}
    \caption{\protect A diagram of the detector components. The PMTs are not shown for the slabs or panels. All components are installed on an aluminum tube.}
    \label{fig:demonstratorDiagram}
\end{figure}

\begin{figure}[!htp]
    \includegraphics[width=\columnwidth]{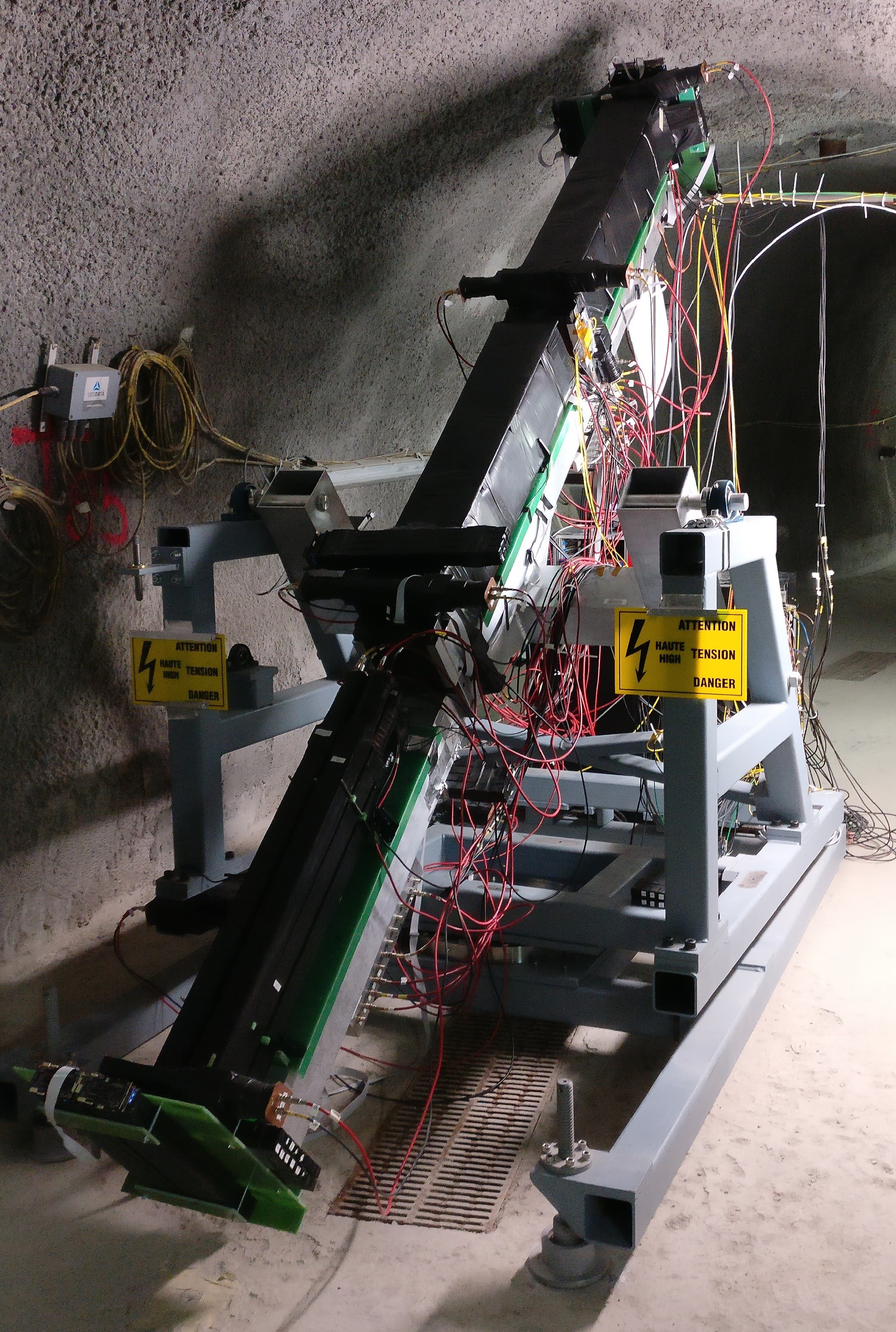}
    \caption{\protect A photograph of the detector, with the panels removed from layer 1 so that the bars may be seen.}
    \label{fig:demonstratorPhoto}
\end{figure}

In addition to the scintillator bars, additional components were installed to reduce or characterize certain types of backgrounds. Lead bricks are placed between the layers to prevent low-energy secondary particles from one
layer from entering another layer. Four scintillator slabs are located along the length of the detector to tag throughgoing particles, provide time information, and shield the bars from neutron radiation. Thin scintillator panels cover the top and sides, providing the ability to reject cosmic muons. Lastly, hodoscopes consisting of $2 \times 2 \times 45$ cm scintillator volumes are used to identify the tracks of beam and cosmic muons. A diagram of the detector components is shown in Fig.~\ref{fig:demonstratorDiagram}, and a photograph of the installed detector is shown in Fig.~\ref{fig:demonstratorPhoto}. All scintillator volumes are comprised of Eljen EJ-200~\cite{Eljen}.

Scintillator light in the bars, panels, and slabs is detected by photomultiplier tubes (PMTs) coupled to the scintillator volumes.
Three types of PMTs are used in order to test different manufacturers and gains: the Hamamatsu R878, Hamamatsu R7725~\cite{Hamamatsu}, and Electron Tube 9814B~\cite{ElectronTubes}.
Analog signals from each PMT are sent to two 16-channel CAEN V1743 digitizers~\cite{CAENV1743}, operating at $1.6\times10^{9}$ samples per second with 12-bit resolution, providing 1024 samples within a 640 ns acquisition window. Each scintillator volume with a PMT attached is referred to as a ``channel''.
Since millicharged particles produced at the CMS IP would traverse the full length of the detector in $\mathcal{O}(10)$~ns, tight timing requirements allow for a significant reduction in combinatoric backgrounds by requiring the coincidence of in time signals in all layers, as described in Section~\ref{sec:search}.
The PMTs are powered by a CAEN SY5527 power supply system~\cite{CAENSY5527}.
The hodoscopes are readout with silicon photomultipliers. Their data are triggered and stored independently from the main data stream.

The data set analyzed for this search was collected during 2018, including periods with the LHC beam providing pp collisions (the ``beam-on'' data set) and periods with no collisions (the ``beam-off'' data set). The beam-off data set provides a statistically independent sample to study background processes. 
The detector is located in an area in the far fringe field of the CMS superconducting magnet, and both beam-on and beam-off data sets were recorded with the magnet at its nominal strength of 3.8~T. The actual magnetic field in the milliQan cavern is measured by magnetic field sensors installed in various positions around the detector. We find this magentic field to be under 2~mT when the CMS magnet is at 3.8~T.
Additional data samples were recorded with the field at 0~T for PMT calibration. 

Data is collected by triggering on the coincidence of at least three channels (triple-coincidence) within a window of 100~ns.
A bar contributes to this coincidence if it has a rising edge consistent with a single photoelectron threshold, and the panels and slabs contribute to the coincidence with more stringent thresholds appropriate for identification of muons from cosmic rays (``cosmic muons") and muons from pp interactions (``beam muons").
The total trigger live-times were 1106 and 1042 hours for the beam-on and beam-off data sets, respectively. The average trigger rate was 14.4~Hz. Outside of the beam time, additional specialized runs were taken using single-channel and double-coincidence triggers, with different thresholds and operating voltage settings, in order to collect data samples for calibrations and validation studies.

\section{\label{sec:calibration}Calibration}

We first calibrate the size of the generated pulses in each channel, which requires a measurement of the average size of a pulse from a single photoelectron (SPE) in each PMT, as well as the mean number of photoelectrons \expNpe generated in each channel by a throughgoing muon. The former is strictly a property of each PMT, while the latter depends on each scintillator, its wrapping, its coupling to the PMT, and the PMT quantum efficiency.

The SPE calibration is performed in-situ by isolating pulses from late-arriving scintillation photons, which largely produce SPEs in the PMTs. The mean SPE area is then found by locating the peak of the resulting pulse area distribution. These measurements are cross-checked by bench measurements of 
SPE waveforms generated by a flashing LED, following the method outlined in Ref.~\cite{Saldanha:pmt}.
In the following, the \Npe of a given pulse is defined
as the pulse area divided by this per-channel SPE calibration, and represents an estimate of the true number of photoelectrons that generated the pulse.

For panels and slabs, the \expNpe calibration is performed directly based on throughgoing beam muons (slabs) and cosmic muons (panels). The measured average pulse area is scaled by the per-channel SPE measurement to calculate the mean number of photoelectrons generated by a beam or cosmic muon. 

For the bars, a direct calibration is not possible as both beam and cosmic muons saturate the readout. Instead, we take an indirect approach, using the fact that the PMT response scales as a power law over a wide range of operating voltages. First, the mean areas of cosmic muon pulses are measured at 5--6 operating voltages, which are low enough to ensure that the PMT signals do not saturate. A power law function is then fit to these points, and extrapolated to the nominal operating voltage. Finally, this number is scaled by the per-channel SPE measurement to arrive at an estimate of the number of photoelectrons generated by a cosmic muon. The validity of the power law assumption is confirmed by separately fitting a power law function to the mean areas of SPE pulses over a range of voltages near the nominal operating voltage. The fitted exponent is found to be consistent with that from the fit to the cosmic muon pulse areas for all bars. The calibrated value of \expNpe for a beam muon traversing the full 80~cm length of a bar varies from 22\,000 to 82\,000; this means that $\expNpe=1$ is expected in the bars for particles of charge $Q/e\sim0.004$--0.007.

The dominant source of uncertainty in each bar's 
\expNpe measurement is the statistical uncertainty from the power law fit and extrapolation, which is 10--20\% depending on the channel, and is uncorrelated between channels. Smaller uncertainties, generally on the order of a few percent, come from differences in the residual magnetic field between calibration runs and data-taking runs; the effect of a low-pass filter applied to the waveforms; and differences between the in-situ and lab-based SPE measurements. These are correlated between bars with PMTs of the same type. 

\begin{figure}[!htp]
    \centering 
    \includegraphics[width=\columnwidth]{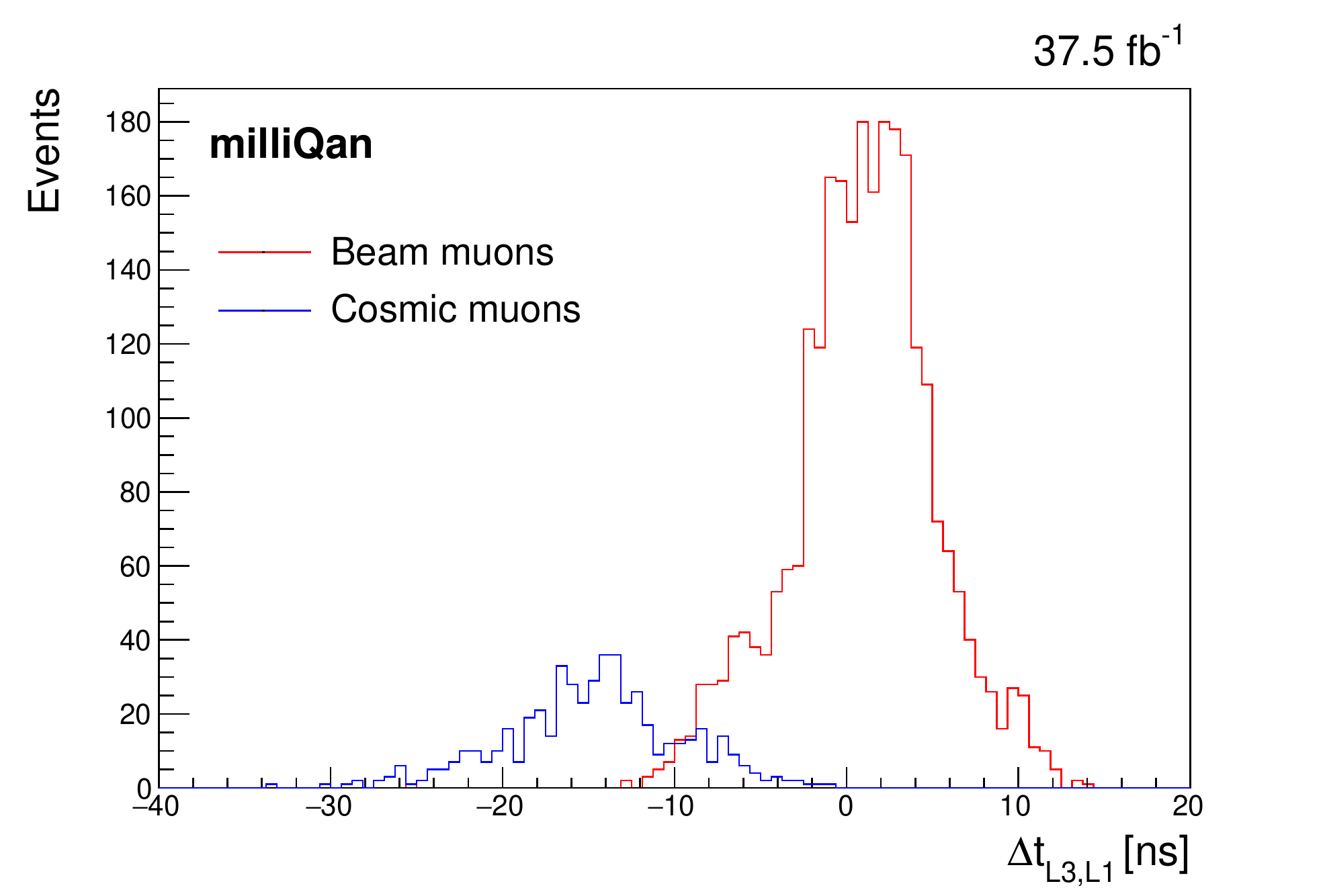}
    \caption{\protect Time difference, measured in a combination of the beam-on and beam-off data sets, between layer 3 and layer 1 for beam (red) and cosmic (blue) muons that travel through the detector.}
    \label{fig:deltaPlot}
\end{figure}

The timing of the PMTs must also be calibrated. The calibration procedure is designed such that a particle traveling near the speed of light through the detector
from the CMS IP should have the same time
value in every bar, panel and slab. This calibration is performed using both beam and cosmic muons. Figure~\ref{fig:deltaPlot} shows the time difference between a muon pulse in layer 3 compared to a muon pulse in layer 1, where the events have been categorized as either beam or cosmic muons based on the timing of the pulses in the slabs. The resolution in the time difference between layers is approximately 4~ns for beam muons which travel through the detector from the CMS IP. An additional correction is applied to account for the dependence of the timing of the pulses on their size. This correction is derived using secondary particles that result from the interactions of beam muons with the detector as they traverse it. The timing resolution degrades as the size of the pulses gets smaller; the resolution of the lowest \Npe pulses passing the selection outlined in Section~\ref{sec:search} is ${\sim}15$~ns.  The modeling of the timing of these secondaries is used to derive a systematic uncertainty in the timing in simulation. 

Using dedicated runs, the trigger efficiency is measured as a function of \Npe for each bar. This information is used to reweight the simulated samples of signals from millicharged particles described in Section~\ref{sec:sim}. The trigger efficiency is measured to reach 100\% for $\Npe > {\sim} 2$ for all channels.  

 \section{\label{sec:mc}Simulation}

\subsection{\label{sec:evgen} Event generation}

We generate pair-production of millicharged particles 
in 13 TeV pp collisions through the Drell-Yan process, as well as from $\Upsilon$, J/$\psi$, $\psi$(2S), $\phi$, $\rho$, and $\omega$ decays into $\mCPp\mCPm$, and from Dalitz decays of $\pi^0$, $\eta$, $\eta'$, and $\omega$.

Drell-Yan events are generated assuming that $\mCP$s are isospin singlet fermions,
using the Lagrangian of Ref.~\cite{Haas:2014dda} implemented in the 
\textsc{MadGraph5}\_a\textsc{mc@nlo}~\cite{Alwall:2014hca} event generator, with a
minimum invariant mass requirement on the millicharged pair of 2 GeV/c$^2$.
Because of the limited integrated luminosity
and the small size
of the demonstrator, there is essentially no sensitivity to Drell-Yan production of millicharged particles in the data set discussed in this paper.

The production cross section and transverse momentum 
\pt distribution of J/$\psi$ and $\psi$(2S) 
in the central rapidity $y$ region is taken from 
calculations of charmonium production
from direct processes~\cite{Ma:2010jj, Ma:2010yw, Ma:2014mri}
and from bottom hadron
decays~\cite{Cacciari:2015fta,Cacciari:2012ny,Cacciari:2001td},
including theoretical uncertainties. Theoretical
calculations of bottomonium production~\cite{Han:2014kxa} are 
not reliable at low transverse momentum ($\pt<15$ GeV)~\cite{Priv_Comml}, 
where most of the cross section lies. As a result,
for $\pt > 20$ GeV we use the cross sections and \pt 
spectra measured at a center-of-mass energy of 13 TeV~\cite{Sirunyan:2017qdw};
at lower \pt we use measurements from 7 TeV 
data~\cite{Aad:2011xv,Aad:2012dlq}, rescaled using
the measured ratio of 13 to 7~TeV cross sections at 
slightly higher rapidity ($2<y<2.5$)~\cite{Aaij:2018pfp}.

All relevant light flavor mesons except $\phi$ mesons are generated
with the minimum bias \textsc{pythia8} generator~\cite{Sjostrand:2007gs}
with the Monash 2013 tune~\cite{Skands:2014pea}.
This is the tune that gives the 
best agreement with several measurements of light meson 
rates and \pt spectra at the LHC, albeit in most cases 
at center-of-mass energies lower than 13~TeV~\cite{Acharya:2017tlv,Acharya:2018qnp,ALICE-PUBLIC-2018-004,Sirunyan:2017zmn}. The Monte Carlo spectra for 
$\eta~(\rho, \omega)$ with $\pt<3$~(1)~GeV are scaled 
down by factors as large as two, based on the 
experimental results cited above.
On the other hand, the production of $\phi$ mesons 
is modeled with the minimum bias
\textsc{pythia6} generator~\cite{Sjostrand:2006za} with the
DW tune~\cite{Albrow:2006rt}, since this Monte 
Carlo setup best reproduces $\phi$ meson data~\cite{Aad:2014rca}.
All \textsc{pythia} Monte Carlo generations are normalized 
to a minimum bias cross section of $80 \pm 10$~mb based on a measurement by ATLAS~\cite{ATLAS_ppxsec}, with an uncertainty taken to cover the difference with respect to a similar CMS measurement~\cite{CMS_ppxsec}.
We assess an additional 30\% uncertainty in the overall 
rate of each process to account for remaining differences between experimental measurements and \textsc{pythia} predictions of the rates of light mesons per minimum bias event, based on the references cited above.

The branching fractions for all vector meson decays, $V \to \mCPp \mCPm$,
as a function of the \mCP mass are calculated using the Van Royen-Weisskopf
formula~\cite{VanRoyen:1967nq}, normalized to the PDG value of the branching fraction for $V \to e^+e^-$~\cite{PhysRevD.98.030001}, and rescaled appropriately for the assumed charge of the \mCP.
The branching fractions
for meson Dalitz decays, 
e.g., $\eta' \to \mCPp \mCPm \gamma$
or $\omega \to \mCPp \mCPm \pi^0$,
as well as the $\mCPp\mCPm$ invariant mass distributions in these decays
are modeled as a function of the \mCP mass and charge
using the partial width for decays into photons, e.g.,
$\eta' \to \gamma \gamma$ or 
$\omega \to \pi^0 \gamma $~\cite{Landsberg:1986fd}, assuming a Vector Dominance Model for the form factors.

\begin{figure}[!htp]
 \includegraphics[width=\columnwidth]{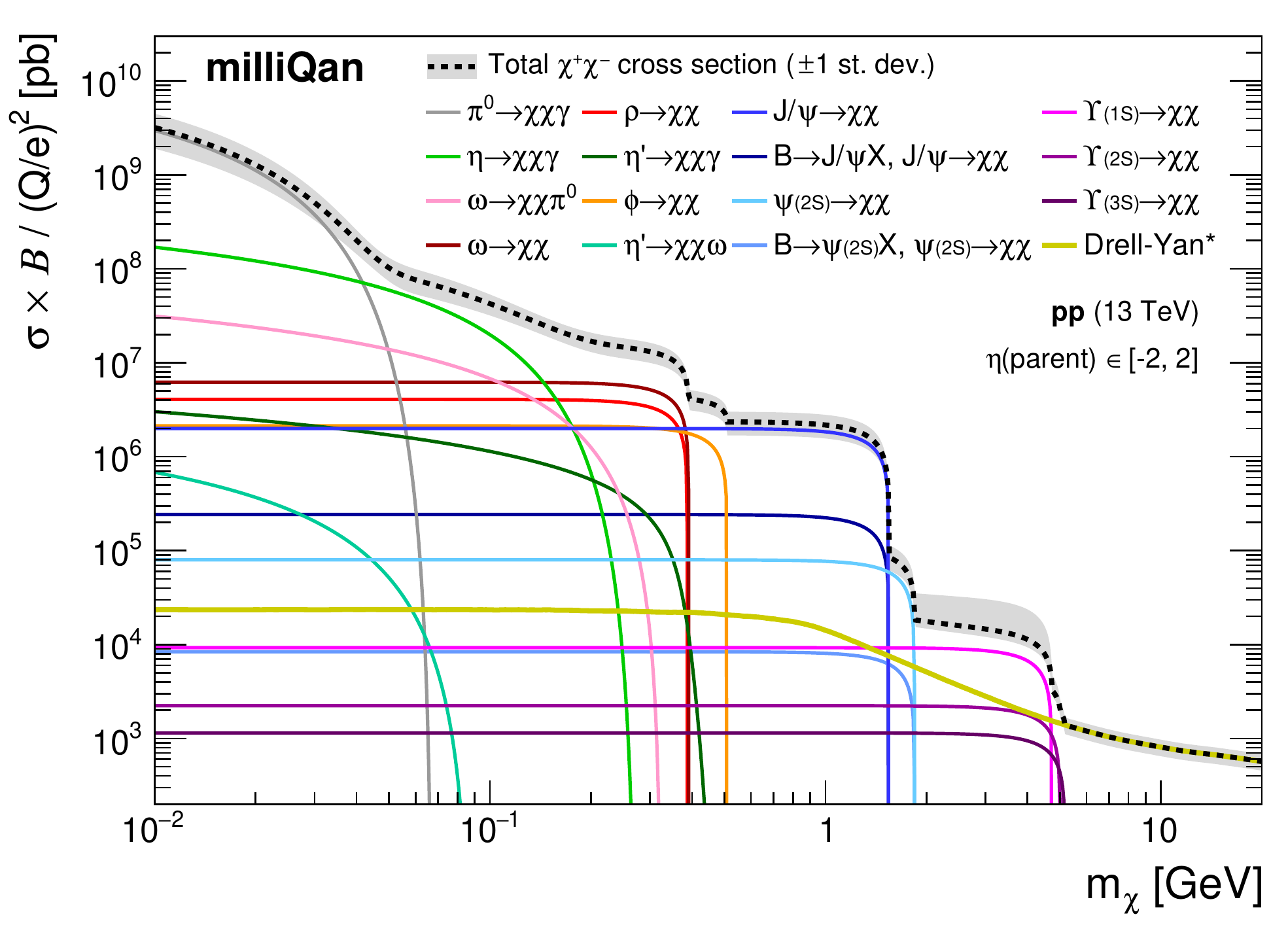}
  \caption{\protect Cross section times branching ratios for production of different particles with $|\eta| < 2$ decaying into $\mCPp\mCPm$ pairs, as a function of \mCP mass. For Drell-Yan the plotted cross section requires at least one of the $\mCP$s to  have $|\eta|<1$.}
    \label{fig:total-rates}
  \end{figure}

Production cross sections of millicharged 
particles from different processes are summarized
in Fig.~\ref{fig:total-rates}.
The possible contribution from millicharged particle production in EM showers in the CMS calorimeters generated by particles from pp collisions is not considered in this analysis.

We also generate inclusive muon events ($\text{pp}\to\text{X}\to\mu$) that are used for
calibration and background studies.
The same theoretical calculation used to obtain the differential cross sections of
J/$\psi$ and $\psi$(2S) from bottom hadron decays is also 
used to generate muons from bottom and charm hadron decays, while 
muons from decays in flight of pions and kaons are generated with \textsc{pythia8} using the Monash 2013 tune. Muons from W and Z decays are taken from \textsc{MadGraph5}, though these electroweak processes contribute only $\sim$3\% of the total muon flux.
Finally, muons are generated using an appropriate angular distribution to simulate cosmic ray events. This angular distribution is derived by assuming a $\cos^2(\theta_{\rm{zenith}})$ distribution at the surface and propagating the muons to the demonstrator using the method described in Section~\ref{sec:sim}.

\subsection{\label{sec:sim} Detector response}

Generated particles are propagated through a simplified model of the material in the
CMS detector, including the magnetic field, and the 17 m of rock
between the CMS cavern and the demonstrator.

\begin{figure}[t]
  \centering
  \includegraphics[width=\columnwidth]{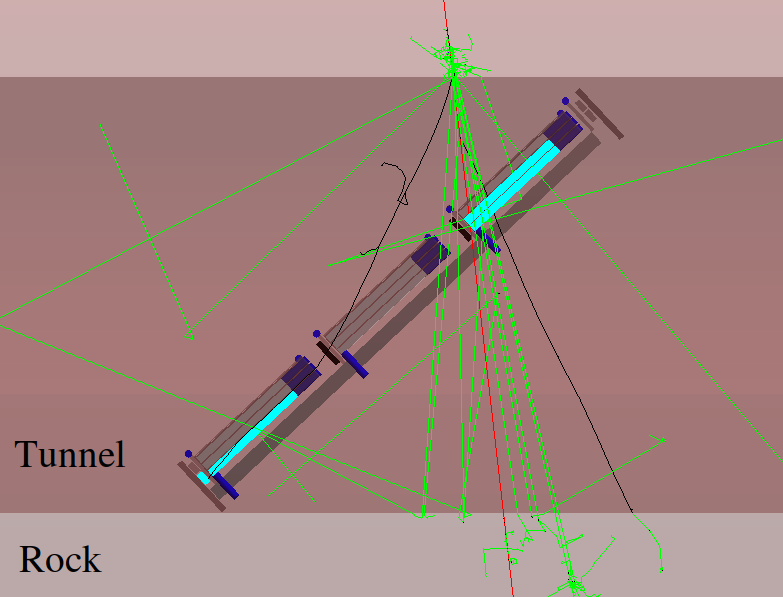}
  \caption{Simulation of the milliQan demonstrator response to a simulated cosmic ray shower event. The incident muon (red) interacts with the cavern walls to produce a shower. Electrons (black) and gamma rays (green) generated in the rock are a significant background source. The interactions of muons and shower particles with detector material produce scintillation photons (cyan).}
  \label{fig:detectorMuon}
\end{figure}

Propagation is performed with fourth-order Runge-Kutta integration, incorporating the effects of the magnetic field, multiple scattering, and energy loss. Particles are propagated until 2 m before the face of the demonstrator, after which they are fed into a full \textsc{Geant4}~\cite{AGOSTINELLI2003250} simulation
of the remaining rock, the drainage gallery where the demonstrator is located, and the demonstrator itself. Similarly, muons from cosmic ray showers are propagated from the surface of the earth to a plane 1~m above the top of the cavern. Muon interactions in the cavern walls generate showers of gamma rays and electrons that significantly contribute to the cosmic ray background. A simulated cosmic ray shower event is shown in Fig.~\ref{fig:detectorMuon}.

The simulation parameters are nominally configured to be consistent with EJ-200 scintillator and existing measurements for Tyvek reflectivity and each PMT species' quantum efficiency~\cite{refl4636942}. Overall photon propagation is handled using \textsc{Geant4}'s UNIFIED optical model~\cite{UNIFIED591410}. Photon propagation is further calibrated by measuring the effects of scintillator roughness and wrapping quality on light attenuation in the scintillator bars and matching the effect in simulation. These measurements indicated a wrapping reflectivity $R=0.97$ and scintillator roughness $\sigma=1\%$, typical values for Tyvek and EJ-200~\cite{Papacz:2010cya}. A per-channel calibration is applied to each PMT quantum efficiency so that the measured \expNpe values agree between data and simulation.
 
The electronic response is simulated using SPE waveform templates measured on a test bench with an LED. SPE templates for each \textsc{Geant4} photoelectron are added together, with appropriate arrival times, and including the individual PMT calibration described in Sec.~\ref{sec:calibration}. The resulting waveform is then added to a randomly selected zero-bias data waveform to properly account for electronic noise. 

\subsection{\label{sec:valid} Validation}
The simulation is validated by studying beam muons as well as cosmic ray events. The absolute rate of beam muons passing through all four slabs is compared with the rate predicted from a simulated sample of muon 
production from heavy-flavor and electroweak decays, as well as
meson decays in flight, as described in Section~\ref{sec:sim}.
We measure a rate of $0.20\pm0.01$ muons/pb$^{-1}$, based on a sample of 7363 muons, in 
agreement with the prediction
of $0.25\pm0.08$ muons/pb$^{-1}$. The dominant uncertainties in this prediction arise from the uncertainty in the 
b$\overline{\text{b}}$ cross section (21\%) and from the 
modeling of the material between the CMS IP and milliQan (25\%). This last uncertainty is derived from a 7\% variation in the total amount of intervening material, which is in turn due primarily to uncertainties in both the thickness and density of the rock layer. This 7\% variation corresponds to a 25\% uncertainty in the muon rate because of the steeply falling muon momentum distribution. This same variation is used to derive a systematic uncertainty in the predicted signal yields, though in that case the effect is much smaller because of the smaller charge of the \mCP.

We additionally perform a comparison of the angular distribution of beam muon trajectories, in order to probe the scattering and magnetic field modeling in simulation and validate the alignment of the detector. We compare rates of muons passing through various subsets of bars that trace a range of angles through the detector. 
The rates of these paths in both the horizontal and vertical directions are measured to be consistent within statistical uncertainties between data and simulation.

\begin{figure}[!htb] 
    \centering \includegraphics[width=1.0\columnwidth]{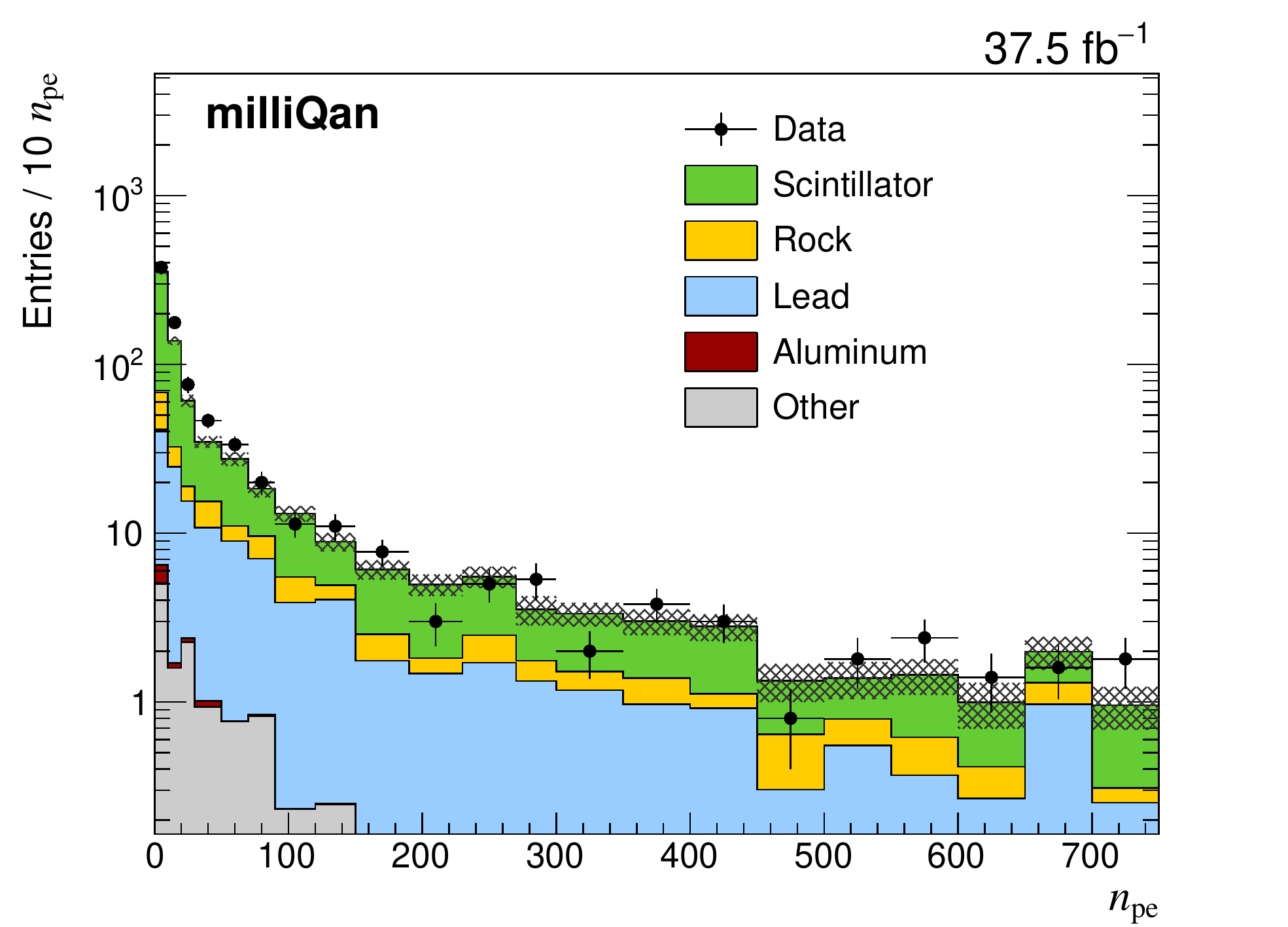}
    \caption{
    \label{fig:npe_nonneighb}
        Comparison of data and simulation \Npe distributions in events with a tagged throughgoing beam muon, for bars that do not contain a pulse consistent with originating from a muon, and are not neighboring any such bars. Note that only 783 out of 7363 tagged muons produced detectable showers entering this figure. Simulation events are categorized based on the material in which the particle(s) that produced the pulse originated. 
        }
\end{figure}

Finally, the \textsc{Geant4} modeling is validated by comparing distributions of photoelectron counts in bars near either beam or cosmic muon trajectories. Hits are expected from electrons and gamma rays produced as the muon travels through the rock or nearby detector material. An example comparison is shown in Fig.~\ref{fig:npe_nonneighb}. Here we show \Npe distributions in data and simulation in events with a tagged beam muon, for bars that do not contain a pulse consistent with originating from a muon ($\Npe<750$), and are not neighboring any such bars. Contributions come primarily from electrons and gamma rays produced in scintillator material, rock, or the lead bricks. Good agreement is seen across a wide range of \Npe levels.

\section{\label{sec:search} Search for millicharged particles}

The search for millicharged particles looks for a signal with the signature of a pulse in each of the three layers of the demonstrator. A number of sources can produce such a signature:

\begin{itemize}
    \item Each PMT has a dark current arising from effects such as the thermal emission of electrons from the cathode. The simplest background source comes from random overlap of three such dark rate pulses. In addition, dark rate counts may overlap with two correlated pulses from another source.
    \item Cosmic muon showers may generate a large number of gamma rays and electrons from an interaction of one or more cosmic ray muons with the rock in the milliQan cavern. This may cause a pulse in each layer of the demonstrator. Such a background could also be expected from a beam muon that travels close to the demonstrator.
    \item Radiation in the cavern, scintillator bars or surrounding material can cause correlated deposits in several bars. The lead blocks placed between layers should reduce the probability of deposit in multiple layers arising from photons or electrons while the slabs provide shielding of neutrons.
    \item Afterpulses, which are small pulses caused by positive ions generated by the ionization
of residual gases in the PMT, can appear several hundred nanoseconds to over a few microseconds after the initial pulse. The afterpulses from correlated deposits in several PMTs may overlap and produce a signal-like signature in the demonstrator. For this to occur, the interaction event that gives rise to the afterpulses must not be triggered because, in this case, the afterpulses will fall in the 125~$\mu$s readout deadtime that follows each triggered event, and will not be recorded.
\end{itemize}

\begin{table*}[ht]

    \topcaption{Sequential impact of each requirement on the number of events passing the selection criteria.}
    \label{tab:sig_regions}
    \resizebox{7in}{!}{%
    \begin{tabular}{c l r r r r r}
    
    \hline
    \hline

    & Selection & Data & Data & Signal & Signal & Signal\\
    & & Beam-on & Beam-off & $m_{\mCP}$ = 0.05 GeV & $ m_{\mCP}$ = 1.0 GeV & $m_{\mCP}$ = 3.0 GeV\\
    & & $t=1106$ h & $t=1042$ h & $Q/e = 0.007$ & $Q/e = 0.02$ & $Q/e = 0.1$\\
    \hline
    
Common & $\geq 1$ hit per layer & 2\,003\,170 & 1\,939\,900 & 136.4 & 34.2 & 5.7\\
Selections & Exactly 1 hit per layer & 714\,991 & 698\,349 & 123.1 & 31.0 & 5.0\\
 & Panel veto & 647\,936 & 632\,494 & 122.5 & 30.8 & 4.9\\
 & First pulse is max & 418\,711 & 409\,296 & 114.3 & 30.6 & 4.8\\
 & Veto early pulses & 301\,979 & 295\,040 & 113.9 & 30.6 & 4.8\\
 & max $\Npe$ / min $\Npe < 10$ & 154\,203 & 150\,949 & 104.2 & 29.6 & 4.7\\
 & $\Delta t_{\rm{max}} < 15$ ns & 5\,284 & 5\,161 & 72.8 & 28.4 & 4.4\\
 & Slab muon veto & 5\,224 & 5\,153 & 72.8 & 28.4 & 4.4\\
 & Straight path & 350 & 361 & 68.4 & 28.1 & 4.2\\
 & $\rm{N}_{\rm{slab}} = 0$ & 332 & 339 & 64.8 & 16.9 & 0.0\\
 & $\rm{N}_{\rm{slab}} \geq 1$ & 18 & 22 & 3.6 & 11.2 & 4.2\\
\hline
SR 1 & $\rm{N}_{\rm{slab}} = 0$ \& min $\Npe \in [2,20]$ & 129 & 131 & 47.4 & 0.4 & 0.0\\
SR 2 & $\rm{N}_{\rm{slab}} = 0$ \& min $\Npe > 20$ & 52 & 45 & 0.0 & 16.5 & 0.0\\
SR 3 & $\rm{N}_{\rm{slab}} = 1$ \& min $\Npe \in [5,30]$ & 8 & 9 & 1.1 & 0.5 & 0.0\\
SR 4 & $\rm{N}_{\rm{slab}} = 1$ \& min $\Npe > 30$ & 4 & 4 & 0.0 & 8.7 & 0.0\\
SR 5 & $\rm{N}_{\rm{slab}} \geq 2$ & 1 & 1 & 0.0 & 2.0 & 4.2\\
        
        \hline
        \hline

    \end{tabular}%
    }
    
\end{table*}

Selections are applied in order to reject contributions from these background sources. If there is a pulse within the acquisition time window in any panel, or in more than one bar in each layer, the event is rejected. These requirements reject backgrounds due to cosmic muon showers, which are expected to cause deposits across the detector. In addition, if there is a pulse in any slab within the acquisition time window consistent with originating from a muon ($\Npe>250$) the event is vetoed. This requirement rejects deposits due to beam and cosmic muons passing close to the bars. The bars with reconstructed pulses are required to be pointing to the CMS IP. This reduces the backgrounds from neutrons, cosmic muon showers and random overlap,
while being efficient for signal, which typically has a small angular spread. To mitigate backgrounds from afterpulses, in each channel a requirement is
made to reject the event if a pulse occurs before the window in which the pulse may be involved in the trigger decision. In addition, the first pulse in each channel must have the largest \Npe value.
Events that contain initial pulses in the bars with a large spread in \Npe (maximum \Npe/minimum \Npe $>10$) are vetoed to reject events containing contributions from different sources, such as dark rate overlap with shower deposits or deposits from two or more shower particles traveling through the demonstrator. Finally, the
maximum calibrated time difference between the first bar pulse in each layer ($\Delta t_{\rm{max}}$)
is required to be less than 15~ns, which is efficient for signals traveling through the detector from the CMS IP and forms a powerful rejection of backgrounds that have different paths through the detector, such as cosmic muon showers, or that have deposits
in each layer that are uncorrelated in their timing, such as dark rate overlap.

Selected events are subsequently categorized into five signal regions (SRs) through requirements on both the number of slabs that contain a pulse and the minimum \Npe of the pulses in the three bars. This categorization allows for sensitivity to a wide range of charge values in the signal parameter space. The definitions of the five SRs are summarized in Table~\ref{tab:sig_regions}. For events with a pulse in each of the three layers the selection criteria provide high efficiency for the targeted models while rejecting the background by more than five orders of magnitude. 

Residual background passing selection is estimated for each signal region by measuring the pass/fail ratio of the timing requirement in events with a hit in each layer, consistent with signal requirements except that the bars do not form a pointing path towards the CMS IP, and then multiplying it by the number of events failing the timing selection that form a pointing path towards the CMS IP. This prediction method relies on the independence of the dominant backgrounds on the pointing path requirement. This method is used rather than taking the prediction from the beam-off data set as it is robust against time dependent drifts in the PMT response. However, the beam-off data set provides a statistically independent sample to validate the prediction without contamination from signal. The results of the beam-off prediction for the SR are summarized in Table~\ref{tab:bkgPrednoBeam}. The uncertainty in the prediction reflects the limited statistics in the regions used to make the prediction. The prediction is shown to be in agreement with the observation for all validation regions. The level of agreement between prediction and observation in each validation region is used to derive a systematic uncertainty
in the prediction.

\begin{table}[ht!]
\centering
\renewcommand{\arraystretch}{1.2}
\topcaption{Summary of the results of the validation using the beam-off data set. The systematic values are derived from the level of agreement between the prediction and observation.}
\resizebox{\columnwidth}{!}{%
\begin{tabular}{cccccc}
\hline
\hline
Region&$\rm{N}_{\rm{slab}}$&min \Npe&Prediction&Observation& Systematic\\
\hline
1&0&[2,20]&$121.2_{-5.9}^{+6.0}$&131&8\%\\
2&0&$>20$&$47.4_{-4.8}^{+5.2}$&45&5\%\\
3&1&[5,30]&$7.8_{-1.8}^{+2.5}$&9&15\%\\
4&1&$>30$&$2.7_{-1.1}^{+2.1}$&4&48\%\\
5&$\geq2$&-&$0.8_{-0.4}^{+1.4}$&1&25\%\\
\hline
\hline
\label{tab:bkgPrednoBeam}
\end{tabular}%
}
\end{table}

\begin{table}[ht]
\renewcommand{\arraystretch}{1.2}
\centering
\topcaption{Summary of the results of the signal region prediction.}
\resizebox{\columnwidth}{!}{%
\begin{tabular}{ccccc}
\hline
\hline
Region&$\rm{N}_{\rm{slab}}$&min \Npe&Prediction&Observation\\
\hline
1&0&[2,20]&$124_{-11}^{+11}$&129\\
2&0&$>20$&$49.9_{-5.4}^{+6.0}$&52\\
3&1&[5,30]&$10.7_{-2.6}^{+3.6}$&8\\
4&1&$>30$&$2.4_{-1.1}^{+2.1}$&4\\
5&$\geq2$&-&$0.0_{-0.0}^{+0.9}$&1\\
\hline
\hline
\label{tab:bkgPredbeam}
\end{tabular}%
}
\end{table}

Given the validation of the background prediction method with the beam-off data set, the SR prediction is made using the beam-on data set. The background contribution from beam processes is estimated from simulation to be less than 2\% for all regions. Results are given in Table~\ref{tab:bkgPredbeam}. The predictions are seen to be consistent with those from the beam-off data set (taking the 6\% difference in collection time into account), which provides additional confidence that the beam-based backgrounds are negligible. The uncertainty in the prediction reflects both the limited statistical power of the regions used for the prediction as well as the systematic uncertainty derived from the validation using the beam-off data set. The predictions are found to be consistent with the observations in all SRs.

\section{Interpretation}

The search is interpreted using the signal model described in Section~\ref{sec:mc}. The full set of selection criteria described in Section~\ref{sec:search} is applied to each event. The efficiency to pass these criteria for three benchmark \mCP masses and charges is shown in Table~\ref{tab:sig_regions}.

There are several sources of systematic uncertainty in the number of signal events entering the SRs. These are evaluated independently for each model point and are summarized below

\begin{itemize}
\item Signal cross section (described in Sec.~\ref{sec:evgen}): typically 15--30\% depending on the mass of the \mCP.
\item Material interactions (described in Section~\ref{sec:valid}: typically 1--5\% depending on the charge and mass of the \mCP. 
\item Pulse timing (described in Section~\ref{sec:calibration}): typically 1--40\% depending on the charge of the \mCP.
\item \expNpe calibration (described in Section~\ref{sec:calibration}): 1--50\% depending on the SR populated by the \mCP. 
\item Limited simulated sample size: up to 30\%.
\end{itemize}

\begin{figure}[!htp]
    \centering 
    \includegraphics[width=\columnwidth]{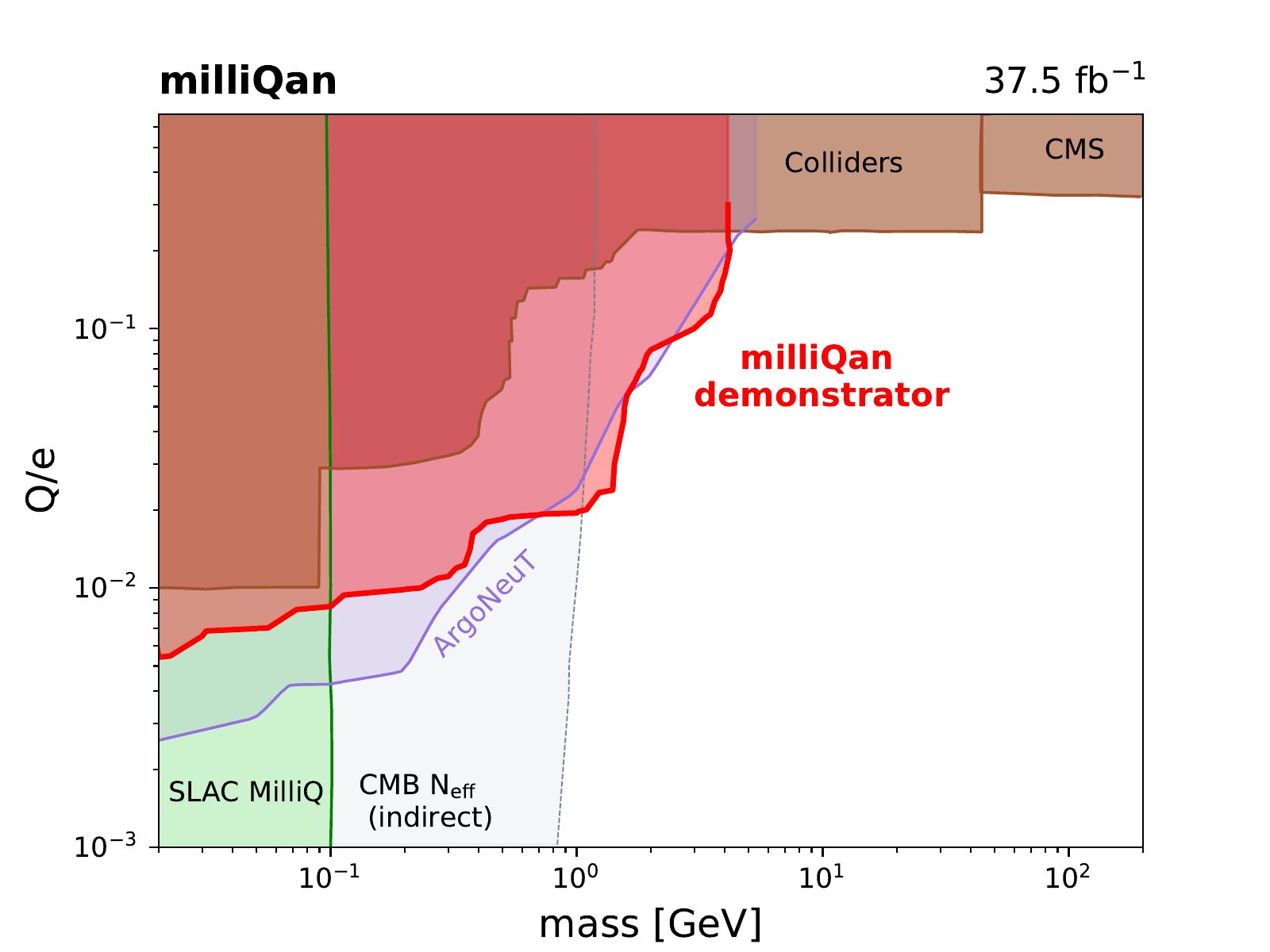}
    \caption{\protect Exclusion at 95\% confidence level compared to existing constraints from colliders, CMS, ArgoNeuT and SLAC MilliQ ~\cite{MilliQ,Vogel_2014,essig2013dark,Chatrchyan_2013,Chatrchyan_2013_2,Acciarri_2020, Davidson:2000hf, Badertscher:2006fm} as well as the indirect constraint from the CMB relativistic degrees of freedom~\cite{Brust:2013ova}.}
    \label{fig:limit}
\end{figure}

Under the signal plus background hypothesis, a modified frequentist approach is used
to determine observed upper limits at 95\% confidence level
on the cross section ($\sigma$) to produce a pair of $\mCP$s, as a function of mass and charge. The approach uses the LHC-style profile likelihood ratio
as the test statistic~\cite{CMS-NOTE-2011-005} and the CLs
criterion~\cite{junk, CLsTechnique}. The observed upper limits are evaluated
through the use of asymptotic formulae~\cite{Cowan:2010js}. Figure~\ref{fig:limit} shows the exclusion at 95\% confidence level in mass and charge of the \mCP. The exclusion is compared to existing constraints, showing new sensitivity for \mCP masses above 700 MeV.

\section{\label{sec:future}Future plans}

In Refs.~\cite{Haas:2014dda, Ball:2016zrp} we assumed the largest irreducible background to the signal would come from dark-current pulses in the PMTs. From experience gained by operating the demonstrator, we now know that an equally important background comes from correlated effects caused by activity in the scintillator (from effects such as environmental radiation or cosmic muon showers). This realization prompted us to revisit the milliQan design, adding a fourth layer in order to mitigate the contribution from these correlated backgrounds.

We have studied the effect of adding a fourth layer with the demonstrator. The demonstrator has three rather than four layers so backgrounds are determined for three-fold coincidence and then extended to four-fold using an additional pulse in a slab. The results of this study indicate that the contribution from pure dark-current overlap drops to a negligible level for the case of four-fold coincidence, even with the somewhat high-noise PMTs that are used in the demonstrator. The calculations presented in Refs.~\cite{Ball:2016zrp, Haas:2014dda} remain a conservative estimate of the milliQan discovery potential since the background with four layers as measured with the demonstrator is significantly smaller than the estimate used in those simulations.

Given the experience obtained from the demonstrator, we are confident that the proposed full-scale detector will perform as expected provided sufficient funding becomes available.

\section{\label{sec:conclusion} Conclusions}

We have deployed a prototype dedicated detector at the LHC to conduct the first search for elementary particles with charges much smaller than the electron charge at a hadron collider. We analyzed a data sample of proton-proton collisions collected
at $\sqrt{s}=13$ TeV provided by the LHC, corresponding to an integrated luminosity of 37.5 fb$^{-1}$. The existence of new particles with masses between 20 and 4700 MeV is excluded at 95\% confidence level for charges varying between $0.006e$ and $0.3e$, depending on mass. 
New sensitivity is achieved for masses larger than $700$ MeV. The successful operation of the milliQan demonstrator and search carried out have shown the feasibility of a dedicated detector for millicharged particles at the LHC and provided important lessons for the design of the full detector. 

\section*{\label{sec:ack} Acknowledgments}
We congratulate our colleagues in the CERN accelerator departments for the excellent performance of the LHC and thank the technical and administrative staffs at CERN. In addition, we gratefully acknowledge the CMS Collaboration for supporting this endeavor by providing invaluable logistical and technical assistance. We also thank Eder Izaguirre and Itay Yavin for their enduring contributions to this idea. Finally, we acknowledge the following funding agencies who support the investigators that carried out this research in various capacities: FWO (Belgium) under the “Excellence of Science – EOS” – be.h project n. 30820817; DFG and HGF (Germany); Swiss Funding Agencies (Switzerland); STFC (United Kingdom); DOE and NSF (USA); Lebanese University (Lebanon).

\bibliographystyle{JHEP3}
\bibliography{milliqan}
\end{document}